\documentstyle[preprint,aps]{revtex}
\begin{document}
\draft
\title{
On the validity of the linear speed selection mechanism for fronts of the
nonlinear diffusion equation}
\author{R.\ D.\ Benguria and M.\ C.\ Depassier}
\address{        Facultad de F\'\i sica\\
	P. Universidad Cat\'olica de Chile\\
	       Casilla 306, Santiago 22, Chile}

\date{\today}
\maketitle
\begin{abstract}
We consider the problem of the speed selection mechanism for the one
dimensional
nonlinear diffusion equation $u_t = u_{xx} + f(u)$. It has been rigorously
shown
by Aronson and Weinberger that for a wide class of functions $f$, sufficiently
localized initial conditions evolve in time into a monotonic front which
propagates with speed $c^*$ such that $2 \sqrt{f'(0)} \leq c^* < 2 \sqrt{
\sup(f(u)/u)}$. The lower value $c_L = 2 \sqrt{f'(0)}$ is that predicted by the
linear
marginal stability speed selection mechanism. We derive a new lower bound on
the
the speed of the selected front, this bound depends on $f$ and thus enables us
to assess the extent to which the linear marginal selection mechanism is valid.
\end{abstract}


\section{Introduction}

In several problems arising in biology, population dynamics, pulse propagation
in nerves, crystal growth, fluid flow and others, it is found that if
the system is
suddenly made unstable, the subsequent dynamics is characterized by the
propagation of fronts.  The systems for which this phenomenon occurs have
received much attention recently, especially related to the problem of pattern
formation.  A small
perturbation at a localized point grows to eventually cover the whole space. An
important problem to be solved has been the determination of the speed at which
the
front of the pattern moves into the undisturbed regions of the system and the
wavelength of the pattern left behind.  (For a recent and extensive
review of this subject we refer to \cite{CH93} and references therein.)
Several authors \cite{DL83,BBDKL85,VS88,VS89} have formulated criteria
that provide an answer to these questions. These criteria are heuristic
extensions to
higher order equations of rigorous results and heuristic arguments which have
been  developed for the nonlinear diffusion equation
\begin{equation}
u_t = u_{xx} + f(u)
\end{equation}
where $f(u) \in C^1 [0,1]$, $f(0) = f(1) = 0$. In what follows we assume that
$f$ is positive in (0,1). In this case $u=0$ is the unstable fixed point
and $u=1$ is a stable fixed point. Aronson and Weinberger \cite{AW78}
have shown that any positive initial condition $u_0(x) < 1$ for all x,
which decays exponentially or faster at infinity will evolve into a
front propagating with speed $c^*$. This asymptotic speed is the lower speed
for
which equation (1)  has a monotonic front joining the stable state $u=1$ to the
unstable state $u=0$. Moreover,
\begin{equation}
2 \sqrt{f'(0)} \leq c^* < 2 \sqrt{
\sup(f(u)/u)}.
\end{equation}
For the  special case of the Fisher--Kolmogorov equation $f(u) = u - u^3$,
$f'(0) = 1$ and $\sup(f(u)/u) = 1$ so that
 $c^*=2$. In general \cite{CEbook},
 for any concave $f(u)$, $\sup(f(u)/u) = f'(0)$, and $c^*=2 \sqrt{f'(0)}$.
The value $c^*=2$ is the value which had been derived by Kolmogorov,
 Petrovsky and Piskunov
 \cite{KPP37} using an heuristic argument (the linear
marginal stability mechanism) which is equivalent to the conjecture that the
asymptotic speed of the front is that for which a perturbation to the front is
marginally stable in the frame moving with the front speed. Based on the
applicability of this argument for the Fisher Kolmogorov equation and more
generally for concave functions $f$ several authors have developed extensions
of
this argument to higher order equations. These generalizations are purely
heuristic, the only rigorous results available being those of Aronson and
Weinberger. In general however, $\sup(f(u)/u)$ is not $f'(0)$
and equation (2) gives a bound on the selected speed. It is known that for some
choices of $f$, and
explicit examples have been given, $c^*$ is greater than 2.
These cases, referred to as those in which a nonlinear marginal stability
mechanism operates, have been generalized \cite{VS89} for higher order
equations based on
the observation that for the nonlinear diffusion equation
the selected front is that with the steepest decay
to zero.
The
exact point of transition from the linear marginal stability to the nonlinear
regime
has been determined for functions $f$ of
the form $f(u) = \mu u + u^n - u^{2 n-1}$ for which an exact solution for a
monotonic front can be given. It has been shown that for $\mu$
smaller than a critical value the solution corresponds to a nonlinear marginal
stability solution \cite{PT92}.  To the best of our knowledge, the only lower
bound on the
speed that,
for general $f$,  shows that the linear speed is not always preferred, has been
given recently by Berestycki and Nirenberg \cite{BN92}. They show that
\begin{equation}
c^2 \ge 2 \int_0^1 f(u)\, du
\end{equation}
from where it is evident that for sufficiently large $f$ the speed exceeds the
marginal value $c_L$.
The purpose of this work is to give a new bound that enables one to evaluate
the regime of validity of the linear marginal stability
mechanism with increased accuracy.
As shown by Aronson and Weinberger, the asymptotic speed of the front is the
lowest for which there is a monotonic travelling wave solution
$u = q(x-c^*\,t)$ of
equation (1). The selected speed satisfies $q_{zz} + c^* q_z + f(q) = 0$,
$\lim u_{z \rightarrow -\infty} = 1$, $\lim u_{z \rightarrow \infty} = 0$,
where
$z = x - c^* t$.
We find it convenient to work in phase space, where monotonic fronts obey an
equation of an order less than the original equation.
  Since the selected speed corresponds to that of a decreasing monotonic
front, we may consider the dependence of its derivative $dq/dz$ on $q$. Calling
$p(q) = - dq/dz$, where the minus sign is included so that $p$ is positive, we
find that the monotonic fronts are  solutions of
\begin{mathletters}
\begin{equation}
p(q)\, {dp\over dq} - c^*\, p(q) + f(q) = 0,
\end{equation}
with
\begin{equation}
p(0) = 0, \qquad p(1) = 0, \qquad p > 0 \quad  {\rm in}\quad (0,1).
\end{equation}
\end{mathletters}
The bound follows in a simple way from equation (4a).
Let $g$ be any positive function in (0,1) such that $h = - dg/dq > 0$.
Multiplying equation (4a) by $g/p$ and integrating with respect to $q$ we find
that
\begin{equation}
\int_0^1 \left( h\, p + {f(q)\over p}\, g \right) dq = c^* \int_0^1 g\, dq
\end{equation}
where the first term is obtained after integration by parts.
However since  $p,\,h,\, f, {\rm and}\, g$ are positive,
we have that for every fixed $q$
\[
h\,p + {f(q)\, g\over p} \ge 2 \, \sqrt{f\, g\, h}
\]
hence we obtain our main result,
\begin{mathletters}
\begin{equation}
c^*\, \ge 2\, {{\int_0^1 \sqrt{ f\, g\, h}\, dq}\over{\int_0^1
 g\, dq}}
\end{equation}
 where
\begin{equation}
g \ge 0  \quad {\rm and} \qquad h = - g' \ge 0 \quad {\rm in}\quad (0,1).
\end{equation}
\end{mathletters}
That this result yields a better bound than that given by equation (3) can be
seen by choosing $g$ so that $ g\, h = f$, and $g(1) = 0$\cite{BD93P}.

Next we illustrate the use of this bound
by applying it to two explicit forms of
$f$. Since here we wish only to illustrate the use of this bound, we
 take three simple trial functions. As a first trial function choose $g$ so
that $f = h = - g_1'$ and $g_1 (1) = 0$. That is
\[
g_1 (q) = \int_q^1 f(x)\, dx
\]
Then
\begin{equation}
c \ge {4\over 3} {{ \left( \int_0^1 f(q)\, dq \right) ^{3/2}}
\over{\int_0^1 q\, f(q)\,
dq}}.
\end{equation}
As a second trial function we choose $g_2 (q) = 1 - q^s$ and the last trial
function $g_3(q) = \exp(-s x)$.
 Consider first the example given in \cite{BBDKL85}, $f(u) = u (1-u) (1 + a
u)$, with $a>0$. This form falls in the category given above for which an exact
solution may be found. The transition from the regime of validity of the linear
marginal stability mechanism to the regime of nonlinear behavior occurs at
$a=2$. The results obtained for this function are shown in Fig.\
 \ref{fig1}. The dotted
line labelled AW correspond to the upper bound $2\sqrt{\sup(f(u)/u)}$, and the
dotted line labelled BN is the bound obtained from equation (3). It crosses
$c^*
=2$ at larger $a$. The solid line
corresponds to the bound with the trial function  $g_3$ with $s=7$, the short
dashed line corresponds to the bound obtained using $g_2$ with $s=.5$ and the
long dashed line is the bound using $g_1$ calculated form equation (7).
Aronson and Weinberger's criterion shows that linear marginal stability is
valid
for $0 < a < 1$, and our bound indicates that it is not valid for $a > 3.6$.
 As we
said above the exact solution for this case is known, the transition value
from linear to nonlinear marginal stability occurs at $a=2$. Next we apply the
bound to the quartic polynomial $ f = x\, (1 - x) ( 1 + a\, x^2)$.
 In this case the
exact solution is not known and neither is the transition value from the
linear to the nonlinear regime. The results are shown in Fig.\ \ref{fig2},
where we have
used the same labelling and type of line as in Figure 1.
Aronson and Weinberger's criterion guarantees that linear marginal stability
is valid for $0 \le a \le 4$, and of the simple bounds calculated
 here the best shown
in the picture corresponds to that obtained with $g_2$ for $s = .1$ which shows
that linear marginal stability is not valid for $a \ge 10.3$. One could of
course attempt to obtain a sharper estimate by choosing better trial functions,
but this is not our purpose here.

In conclusion, it is evident
from the present results  that for all non concave functions
$f(u)$ the
linear speed $c_L$ is the asymptotic speed in a rather limited region.
 There is no
substantial difference in the behavior of arbitrary polynomials, for which no
exact  solutions are known, with those already analyzed in the literature for
which the exact solution and point of transition can be calculated.
 Once the function $f$ becomes
sufficiently large, the selected speed will be that of the so called nonlinear
front.  Given the limited validity of the linear selection mechanism for the
nonlinear diffusion equation a similar situation can be expected for higher
order equations. Moreover, since the lower bound on the speed depends on the
integral properties of $f$,
 it is not difficult to imagine a situation where two functions
are identical near the origin and differ significantly near $u=1$. In that case
it is possible that the asymptotic speed for one of them be the linear value
and
for the other
the nonlinear value. No local
analysis of the approach to $u=0$ can then predict the transition from the
linear to the nonlinear marginal stability regime. Finally we wish to point out
 that the
analysis of monotonic fronts in phase space is useful not only in the case
presented here but for generalized diffusion equations and in higher order
equations as well.
For the porous media equation
\[
u_t = (u^m)_{xx} + f(u), \qquad m \ge 1
\]
with $f(0) = f(1) = 0$, and $f > 0$ in (0,1) monotonic fronts may exist for
\[
c\, \ge 2\, {{\int_0^1 \sqrt{ f\, \sigma\, h}\, dq}\over{\int_0^1
 \sigma\, dq}}
\]
where $\sigma(u) > 0$ must be chosen so that
\[
 h(u) \equiv - m\, u^{m-1}\, \sigma'(u) > 0 \qquad {\rm in} \quad (0,1).
\]
For a more general equation
\[
u_t = (\phi (u))_{xx} + f(u), \qquad {\rm with}\quad \phi' > 0 \quad {\rm} \in
\quad (0,1), \quad \phi (0) = 0,
\]
with the same conditions on $f$, monotonic fronts may exist for
\[
c\, \ge 2\, {{\int_0^1 \sqrt{ f\, \sigma\, h}\, dq}\over{\int_0^1
 \sigma\, dq}}
\]
where $\sigma(u) > 0$ must be chosen so that
\[
 h(u) \equiv  - \sigma'(u)\,
\phi'(u) > 0.
\]
The details will be given elsewhere.
It has been applied by us to obtain bounds on the speed of certain third order
nonlinear differential equations of the type which arise in crystal growth
problems \cite{BD93}
 and for the dispersive Kuramoto-Sivashinsky equation \cite{BD93P}. It also
enables
one to characterize the type of functions $f(u)$ for which the exact point of
transition to the nonlinear regime in the the nonlinear diffusion equation (1)
 can be calculated without solving the
equation explicitly \cite{BD93P}.

\section{Acknowledgments}

This work has been partially supported by Fondecyt project 1930559.


\vspace{2.0cm}
\begin{figure}
\caption{
Bounds on the speed of the monotonic fronts for the exactly solvable case
$f(u) = u\, (1 -u) (1 + a\,u)$. The dotted lines are the upper and lower bounds
on the speed obtained from equations (2) and (3).
  The solid and dashed lines show bounds
obtained from equation (6) with different trial functions.
 The bound obtained with the
simple trial function $\exp (-7x)$, shown with a solid line, indicates
that linear marginal stability is not valid for $a>3.6$.  The exact value for
the transition is $a=2$.}
\label{fig1}
\end{figure}

\begin{figure}
\caption{
Bounds on the speed of the monotonic fronts for the quartic polynomial
$f(u) = u\,(1-u) (1 + a\,u^2)$ for which the exact solution is not known.
The labelling of curves is as in Figure 1. The bound obtained with the trial
function $1 - u^s$ with s=.1, shown with the short-dash
 line, indicates that linear
marginal stability is not valid for $a > 10.3$. }
\label{fig2}
\end{figure}
\end{document}